\documentclass[useAMS,usenatbib]{mn2e}
\usepackage{natbib}
\usepackage{flushend}
\usepackage{epsfig}
\usepackage{amssymb}
\usepackage{rotating}
\let\la=\lesssim            
\let\ga=\gtrsim
\bibpunct{(}{)}{;}{a}{}{,}

\newcommand{\hypz}{{\it Hyperz}}

\newcommand{\sex}{{\it SExtractor}}
\def\farcs{\hbox{$.\!\!^{\prime\prime}$}}
\def\re{\hbox{$R_{\rm e}$}}
\def\ks{\hbox{$K_{\rm s}$}}
\def\msun{\hbox{$M_\odot$}}
\newcommand{\sbzk}{$sBzK$}
\newcommand{\sbzks}{$sBzK$s}
\newcommand{\pbzk}{$pBzK$}
\newcommand{\pbzks}{$pBzK$s}

\title[]{High-redshift elliptical galaxies: are they (all) really compact?}
\author[C. Mancini , et  al.]{C. Mancini$^{1}$\thanks{E-mail:chiara.mancini@oapd.inaf.it},E. Daddi$^{2}$, A. Renzini$^{1}$, F. Salmi$^{2}$, H. J. McCracken$^{3}$, A. Cimatti$^{4}$,
\newauthor
M. Onodera$^{2}$, M. Salvato$^{5}$, A. M. Koekemoer$^{6}$, H. Aussel$^{2}$, E. Le Floc'h $^{2,7}$, C. Willott$^{8}$\\
$^{1}$Osservatorio Astronomico di Padova (INAF-OAPD), Vicolo dell'Osservatorio 5, I-35122, Padova Italy\\
$^{2}$CEA-Saclay,DSM/DAPNIA/Service d'Astrophysique, 91191 Gif-Sur Yvette
     Cedex, France\\
$^{3}$Institut d'Astrophysique de Paris, UMR7095, Universit\'e Pierre et
	 Marie Curie, 98 bis Boulevard Arago, 75014 Paris, France\\
$^{4}$Dipartimento di Astronomia, Universit\`a 
     di Bologna, via Ranzani 1, I-40127 Bologna, Italy\\
$^{5}$California Institute of Technology, 105-24 Robinson,
1200 E. California Blvd., Pasadena CA 91125, USA\\
$^{6}$Space Telescope Science Institute, 3700 San Martin Drive, Baltimore, MD 21218, USA\\
$^{7}$Institute for Astronomy, Univ. of Hawaii, 2680 Woodlawn Drive, Honolulu, HI 96822, USA\\                    
$^{8}$Herzberg Institute of Astrophysics, National 
Research Council, 5071 West Saanich Rd, Victoria, BC V9E 2E7, Canada}

\begin{document}
\flushend


\pagerange{\pageref{firstpage}--\pageref{lastpage}} \pubyear{2002}

\maketitle

\label{firstpage}

\begin{abstract}
We investigate the properties of 12 ultra-massive passively evolving  
early type galaxies (ETGs) at $z_{\rm phot}>1.4$ in the COSMOS 2 deg$^2$ field. These 12 ETGs were selected as \pbzks, have accurate $1.4\lesssim z_{phot} \lesssim 1.7$,
high S\'ersic index profiles typical of ellipticals, no detection at 24$\mu$m,
resulting in a complete ETG sample at $M*>2.5\times10^{11}M_\odot$ (Chabrier IMF). 
Contrary to previous claims, the half light radii
estimated in very high S/N imaging data from HST+ACS are found to be large for most of the sample, consistent with
local ellipticals.  If the high redshift ETGs with $M*<2.5\times 10^{11}M_{\odot}$  are really
small in size and compact as reported in previous studies, our result may suggest
a ``downsizing" scenario, whereby the most massive ETGs reach their final
structure earlier and faster than lower mass ones. However, simulating galaxies
with morphological properties fixed to those of local ETGs with the same stellar
mass show that the few compact galaxies that we still recover in our sample
can be understood in term of fluctuations due to noise preventing the recovery of the extended low surface  brightness
halos in the light profile. Such halos, typical of S\'ersic profiles, extending even up to 40 kpc,
are indeed seen in our sample. 
\end{abstract}


\section{Introduction}

Understanding the formation processes of galactic spheroids, i.e.,
bulges and early-type galaxies (ETG), remains a central theme in the
context of galaxy formation and evolution. The
discovery of a widespread population of passively evolving ETGs at
$z>1.4$ proved that quenching of star formation in massive galaxies is
well under way by $z\sim 2$
\citep{2004Natur.430..181G,2004Natur.430..184C}.

Further studies of high redshift ETGs then revealed an unexpected
property of a sizable fraction of them: many such ETGs appear to have
much smaller effective radii ($\re$, by factors $\sim 2$ to $\sim 4$)
with respect to ETGs of comparable stellar mass ($M*$) in the local
Universe (Daddi et al. 2005, D05), a result then confirmed by several
other studies \citep[e.g.][]{2006ApJ...650...18T, 2007MNRAS.382..109T,
2007MNRAS.374..614L,
2007ApJ...656...66Z,2008ApJ...682..303M,2007ApJ...671..285T,
2008A&A...482...21C, 2008ApJ...677L...5V,
2008ApJ...687L..61B,2009ApJ...695..101D, 2009MNRAS.392..718S,
2008arXiv0806.4604R}.  ETGs with similar stellar densities appear to
be extremely rare in the local Universe \citep{2009ApJ...692L.118T},
although it has been argued that such compact high redshift ETGs have
survived as the cores of present-day massive spheroids
\citep{2009arXiv0903.2479H}.

It has also been argued that the $\re -M*$ relation for ETGs would
keep evolving also from $z\sim 1$ all the way to $z\sim 0$, with a
steady increase of $\re$ for given $M*$
\citep{2008ApJ...688...48V,2009arXiv0901.1318B}.

The formation of very compact ETGs at high redshift may not be a
problem: submillimetre galaxies appear to have comparable masses (largely in gas form) and
radii, hence may likely be precursors to compact ETGs \citep{2008A&A...482...21C,2008ApJ...680..246T}, turning into them upon expulsion of
their residual gas. However, no generally accepted explanation has yet
been established on how such compact ETGs would evolve into their present
descendants, i.e., how they can inflate to reach 2--4 times larger
effective radii.

ETG-ETG (dry) merging soon appeared to be a plausible mechanism
\citep[e.g.,][]{2006ApJ...648L..21K}, but most such events would
result in an increase of both mass and radius such to move galaxies
parallel to the local $R_{\rm e}-M*$ relation, rather than towards it
(Cimatti et al. 2008; Saracco et al. 2009). Moreover, major dry
merging events may be too rare anyway, as the vast majority of
high-$z$ ETGs do not appear to have close ETG companions of similar
brightness (Cimatti et al. 2008). Having excluded major mergers,
accretion of many satellites (i.e., minor dry mergers) were then
considered, either adding an extended envelope to the compact core
(Cimatti et al. 2008), or expanding such core by gravitationally heating \citep{2007ApJ...658..710N}. 
It has also been suggested that the expansion of high-$z$ ETGs would be the result of AGN feedback
driving the rapid expulsion of the residual gas
\citep{2008ApJ...689L.101F}, but such expansion must take place in at
most a few dynamical times ($\sim 10^8$ yr), hence only very few
objects could be caught as already passive and still compact. 

On the other hand, the possibility that the apparent small radii may
not be real was not completely excluded. This could be either the
effect of part of the light coming from a centrally concentrated
source (e.g., an AGN or central starburst), or to some systematic bias
not having been taken in full account (D05). 
Recently, \citet{2009AJ....137.3942L} have argued for the presence of a radial age
gradient in local ETGs, that would result in an apparent decrease of
$\re$ with increasing redshift, as the younger, more centrally
concentrated population differentially brightens w.r.t. the older and
broader stellar component. However, there appears to be no strong colour
gradients in the few cases in which both optical and near-IR HST
imaging is available for high-$z$ ETGs \citep[e.g.][]{2007ApJ...671..285T},
or within different ACS bands (D05). 

In this letter we present the results of two-dimensional (2D) surface
brightness profile fitting of a complete sample of 12
extremely massive ETGs ($M*>2.5\times 10^{11}M_{\odot}$) at $z\gtrsim
1.4$, using the HST+ACS/WFC images of the COSMOS 2 deg$^2$ field
\citep{2007ApJS..172..196K}. Selection of high-z ETGs over such a
large area allows us to pick the most massive/luminous high-$z$ ETGs,
and tackle the size issue with the highest S/N ratio.

\section[]{Data, sample selection, and SED fitting}

\begin{figure}
\includegraphics[width=8.5cm]{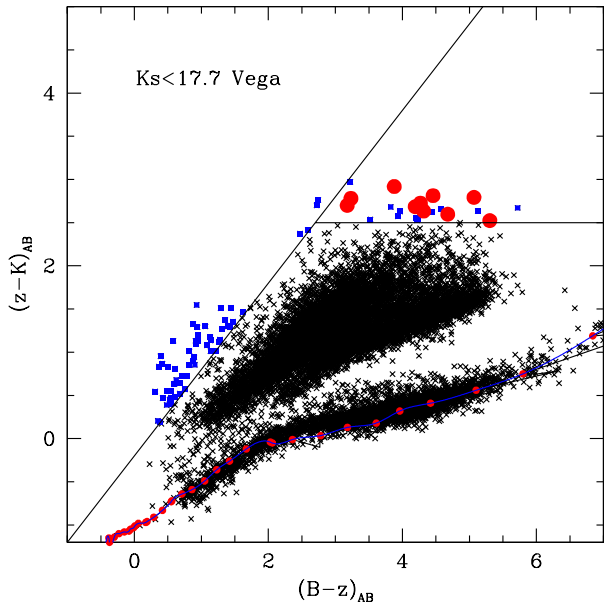}
\caption{The $BzK$ two-colour diagram for all objects with $\ks<17.7$ (Vega system) in the COSMOS field. The 12
bona-fide passive ETGs are shown as red circles. Blue squares with $(z-K)<1.6$ are most likely AGN dominated galaxies, having point-like morphology and/or X-ray emission. The remaining blue squares with redder $(z-K)$ colours show \sbzk\ galaxies and \pbzk\ contaminants rejected for being star-forming. 
  The blue solid line and the red dots in the bottom show the expected colours for the stars \citep[from
  ][]{1997A&AS..125..229L}, overlapping with the locus of real star
  sequences.}
\label{fig:fig1}
\end{figure}

A sample of extremely massive ETGs was extracted from the catalogue of
$K$-selected galaxies in the 2 deg$^2$ COSMOS field (McCracken et
al. 2009). From this catalogue we extracted all the objects with $\ks({\rm Vega})<17.7$, whose corresponding $BzK$ plot is shown in Fig.~\ref{fig:fig1}. From this sample we then selected the
brightest galaxies satisfying the passive BzK galaxy criterion 
of \citet{2004ApJ...600L.127D}, for which the $\ks({\rm Vega})<17.7$ limit roughly corresponds
to $M*> 2.5\times 10^{11}\msun$ at $z\sim1.5$ for a Chabrier IMF. This resulted in a 
sample of 22 sources, having retained only galaxies with IRAC 3.6~$\mu$m detections 
with separations $<0''.6$ from the
$K$-band
positions, a criterion required to avoid objects affected by blending
in the IRAC and MIPS images \citep[see, e.g.][]{2007ApJ...670..156D}. We notice that 4 star-forming $BzK$ galaxies (\sbzk) are found to
$K<17.7$, after excluding likely AGNs among \sbzks, i.e., the blue squares with $(z-K)<1.6$ in Fig.~\ref{fig:fig1}. The colours of these 4 objects are close to the
\pbzk\ boundary and in principle some genuine ETG could have been scattered there just by photometric errors
noise. However, we find that all 4 objects are well detected at $24~\mu$m. 

\begin{figure*}
\includegraphics[width=\columnwidth]{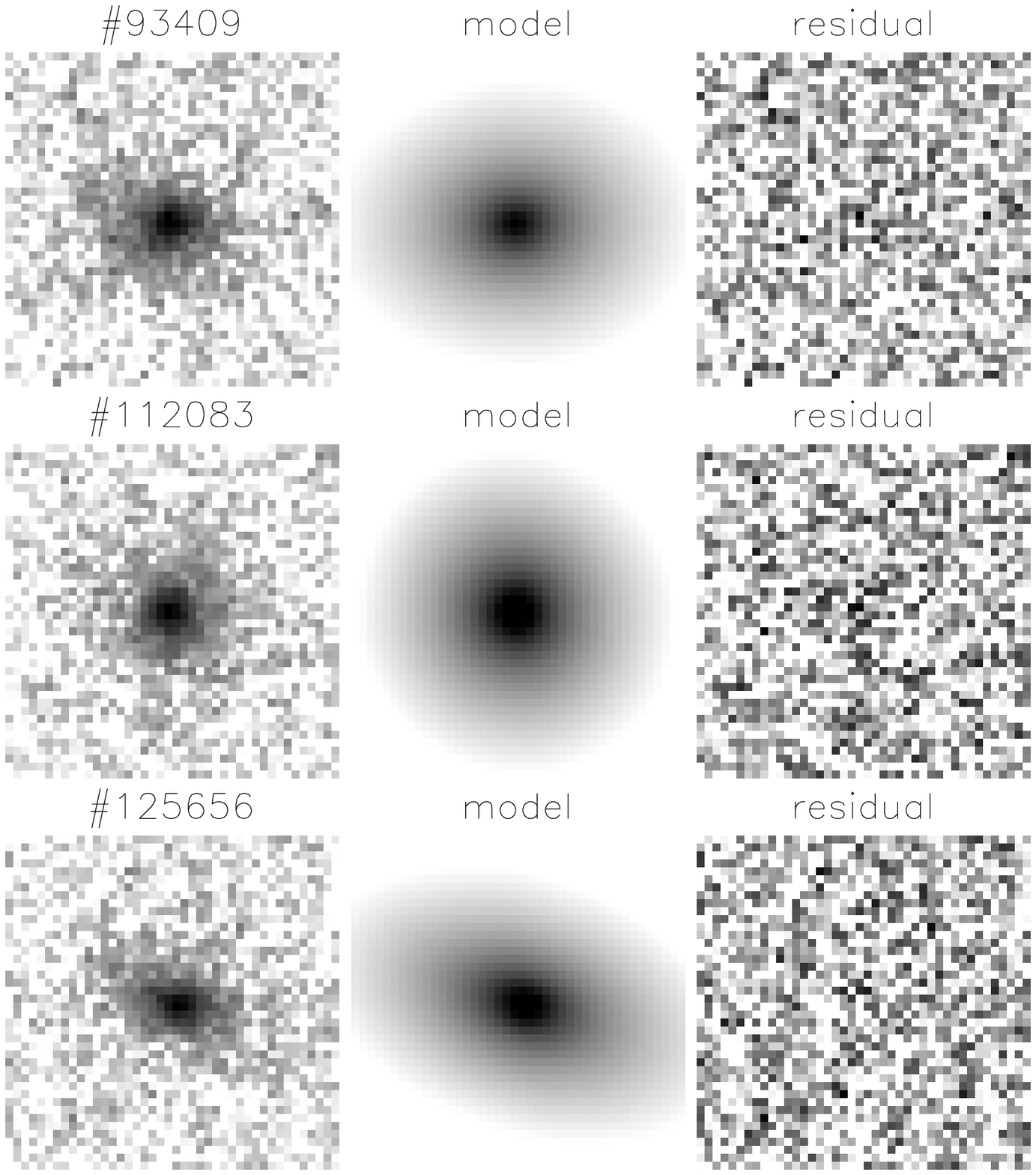}\includegraphics[width=\columnwidth]{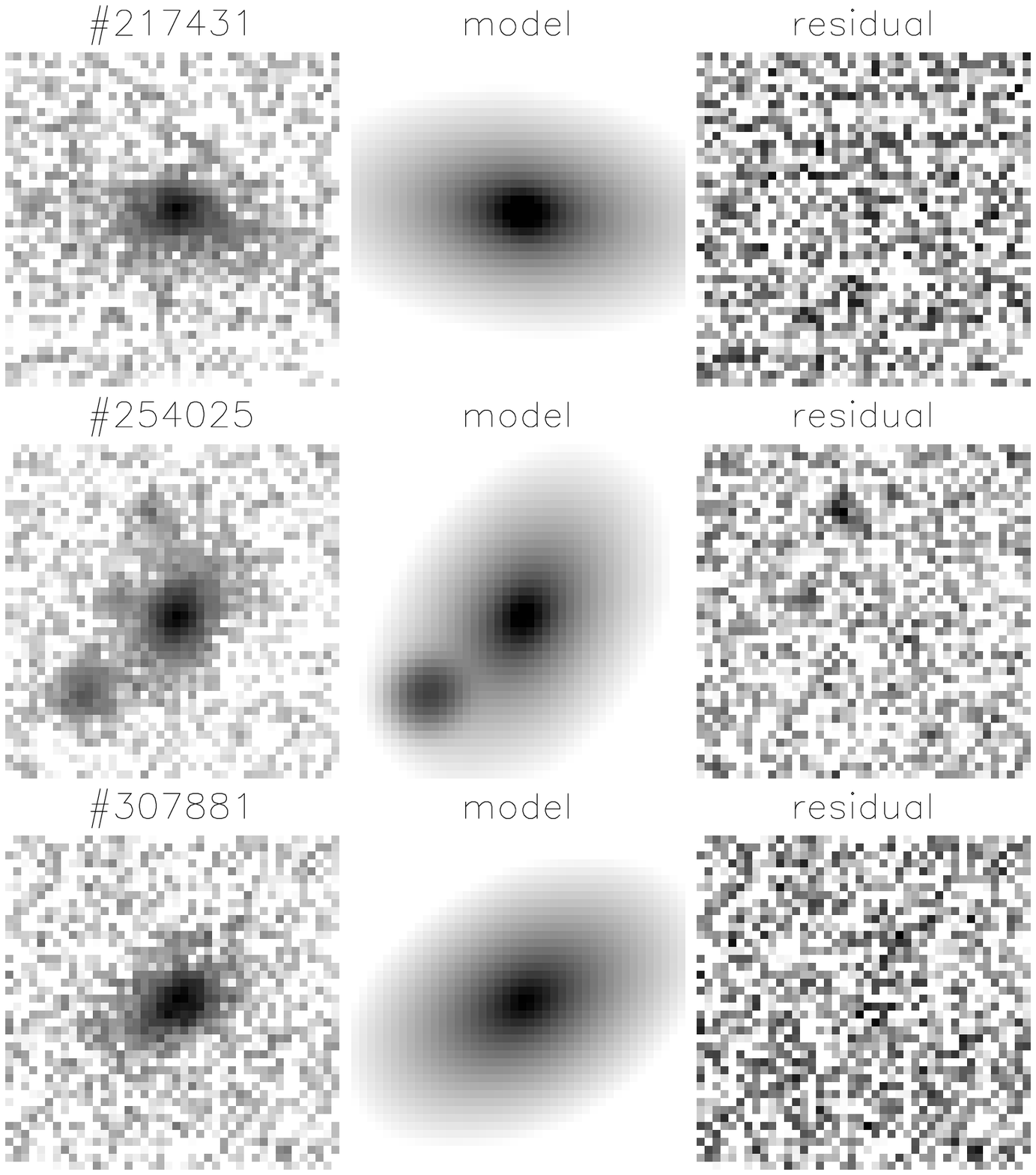}
\includegraphics[width=\columnwidth]{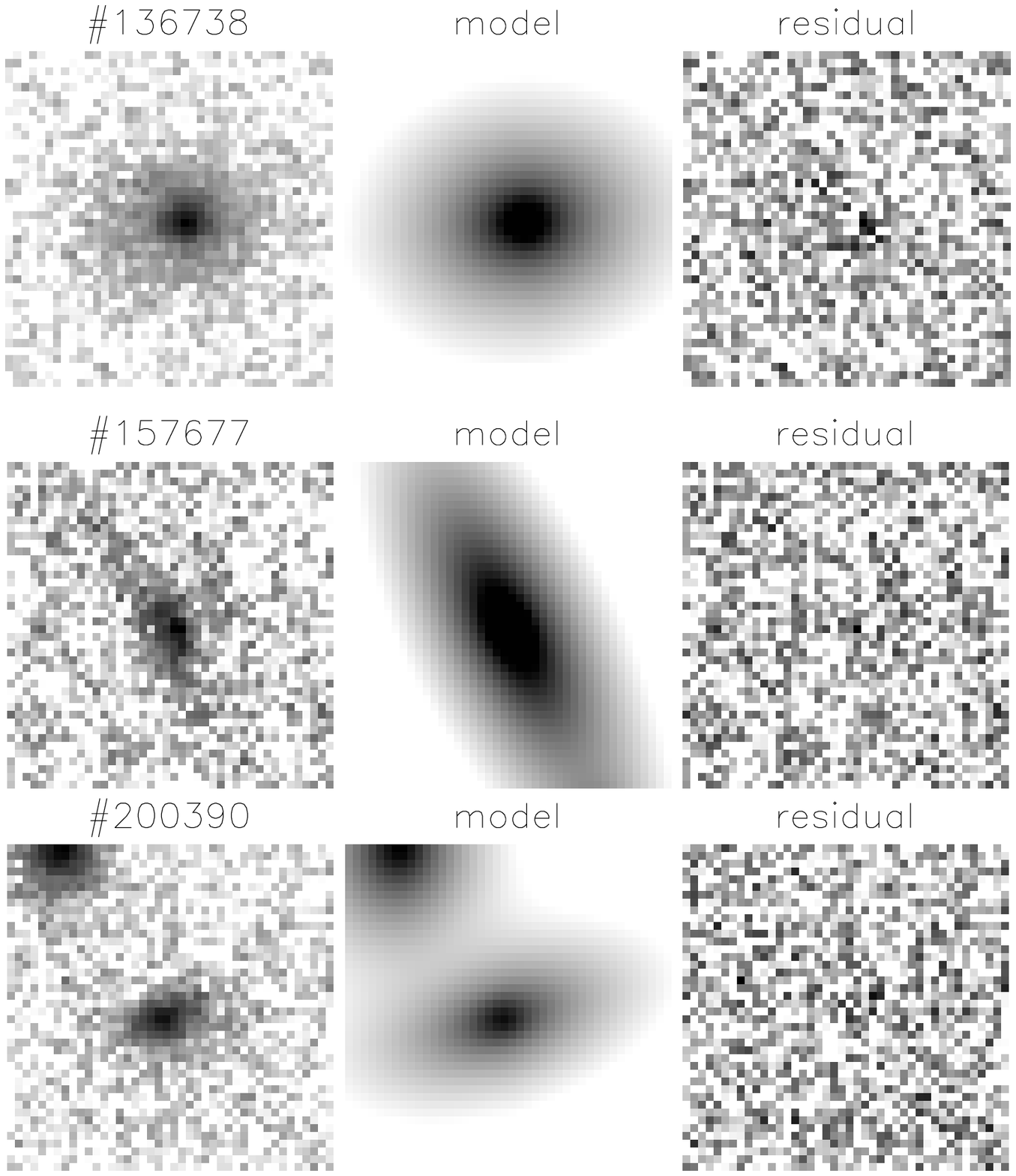}\includegraphics[width=\columnwidth]{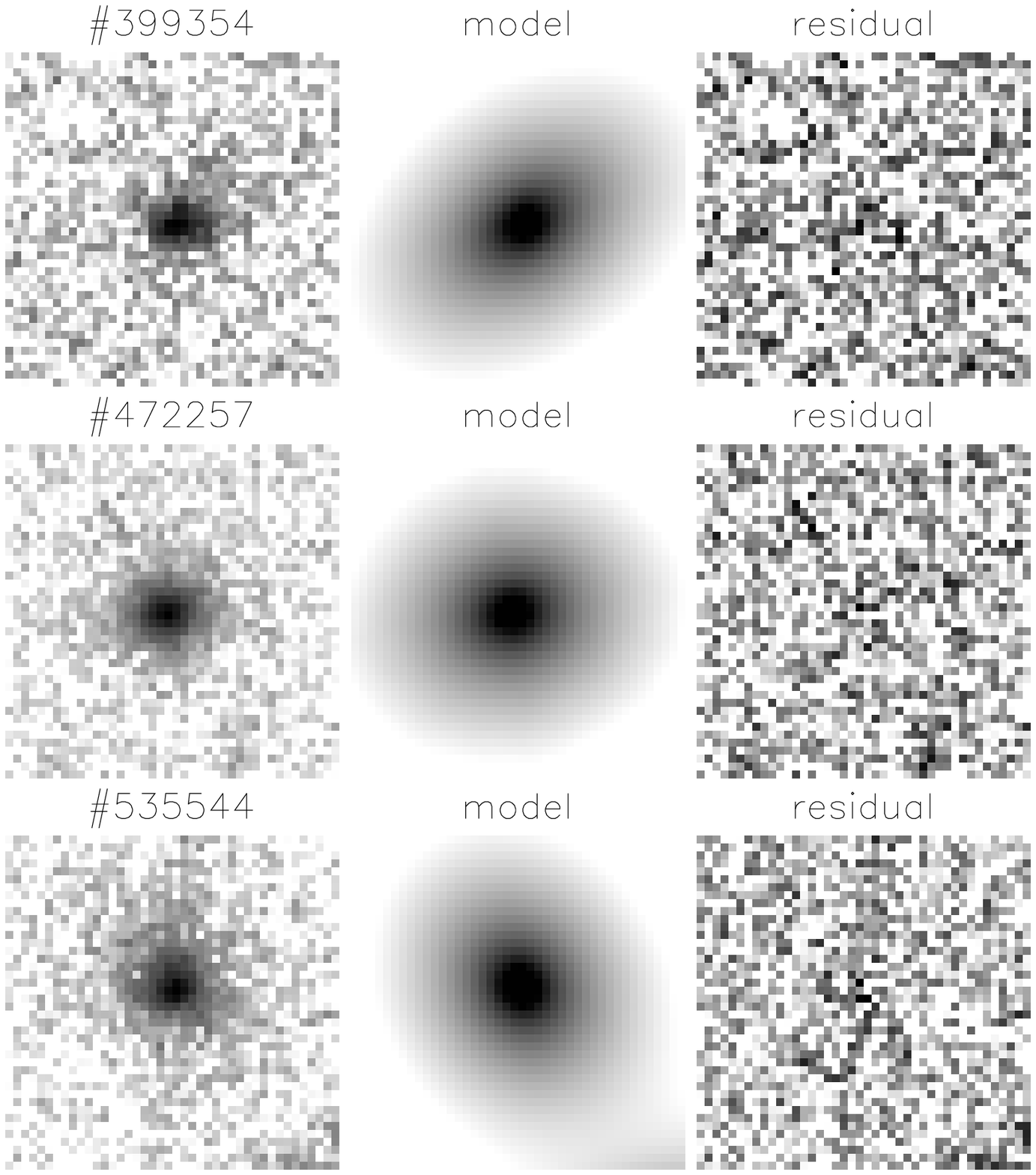}
\caption{ACS $I$-band 2.5~arcsec cutouts for the 12 bona-fide ETGs, their best-fit models and the corresponding residuals. Images 
are displayed in logarithmic scale to enhance low surface brightness details.
 }\label{fig:fig2}
\end{figure*}

\begin{figure*}
\includegraphics[width=\textwidth]{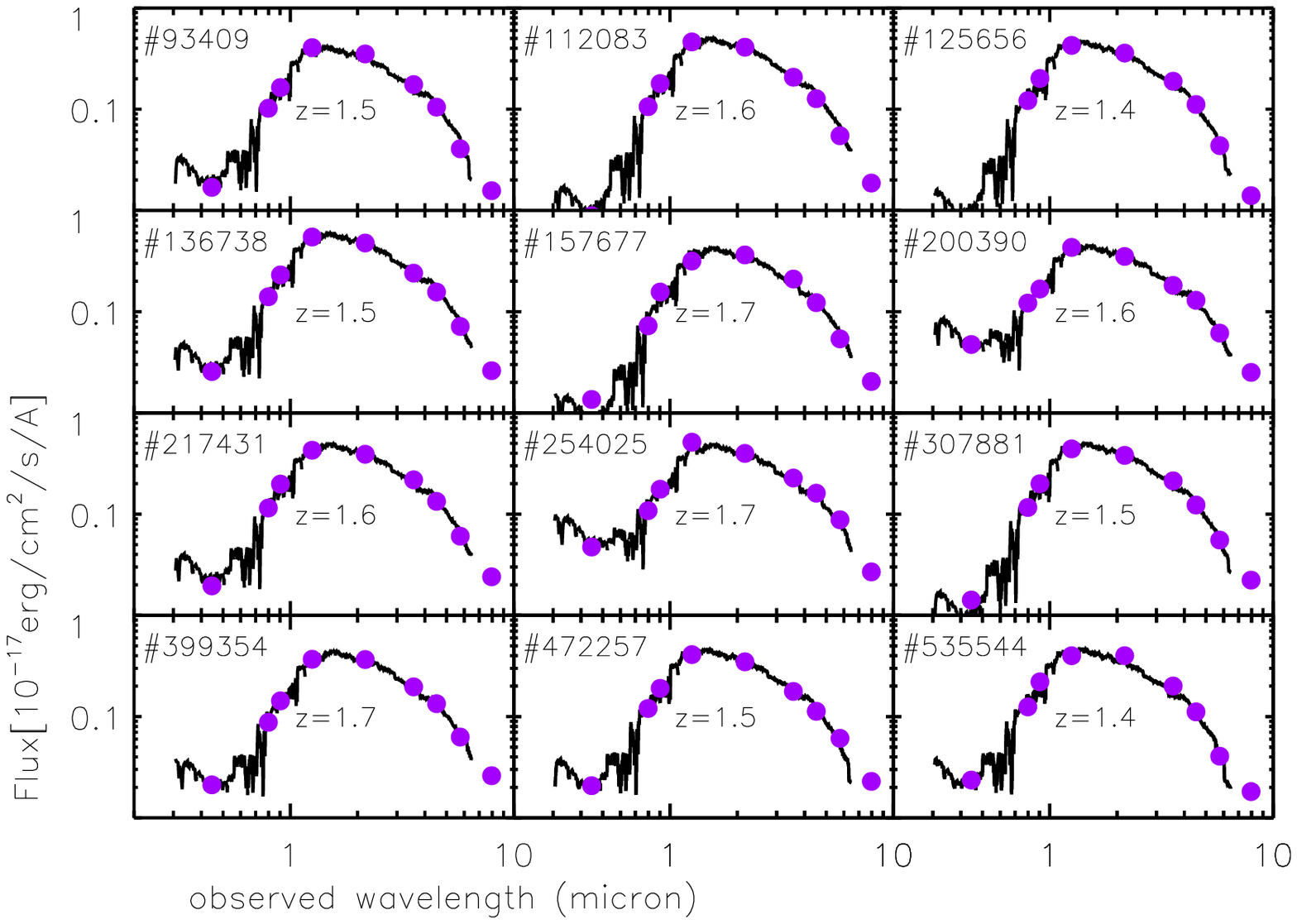}
\caption{Multicolour SED and best-fit SEDs templates (Maraston 2005, Chabrier IMF) of the 12 bona-fide ETGs.}
\label{fig:fig3} 
\end{figure*}

Since these 22 \pbzk-selected objects could include heavily
dust-reddened star-forming galaxies at $z<1.4$, we carefully checked
the properties of the sample to solidly identify the bona-fide ETGs and discard
possible contaminants.
We requested all of the following criteria for retaining ETG candidates:

\noindent
i) non detection at 24~$\mu$m in the {\it Spitzer}+MIPS data
\citep{2007ApJS..172...86S}. We used the most recent deep catalogue by Aussel \& Le Floc'h, reaching 5$\sigma$ completeness levels down to 80$\mu$Jy (Le Floc'h et al. 2009, submitted to ApJ). This corresponds to
a limit of Star Formation Rate (SFR) $<50 M_\odot$~yr$^{-1}$ for $z\sim1.5$, using the models 
of Chary \& Elbaz (2001);

\noindent
ii) elliptical like compact morphology, based on 
a visual inspection of the ACS $I$--band images (Fig.~\ref{fig:fig2});

\noindent
iii) multicolour SEDs best fitted with old, passively evolving
populations with no dust reddening as showing in Fig.~\ref{fig:fig3}. SEDs were constructed
using COSMOS photometry in the $B$-,\ $i$-,\ $z$-,\ $J$,\ $Ks$-band data
consistently with McCracken et al (2009), expanded to include photometry in
the first 3 IRAC bands (from Ilbert, Salvato, \& S-COSMOS collaboration, \citet{2009ApJ...690.1236I}.

We found that 12 of the 22 \pbzk\ galaxies do satisfy all these criteria
and are retained as our sample of bona-fide $z\gtrsim1.4$ ETGs. The
remaining 10 galaxies that were discarded are found to fail multiple
criteria in most cases: they tend to be 24$\mu$m detected (6/10), have
irregular (or point like, unresolved) morphologies in the ACS imaging
(7/10), are best fitted by star forming galaxy templates with large
amounts of reddening (7/10), or show evidence for strong AGN
contamination of the optical/near-IR light (3/10) as indicated by the
`power law' shape of the SED and X-ray and/or radio detection. These
criteria are very efficient in selecting truly passively evolving
ETGs, as confirmed by their spectroscopic identification in the HUDF by D05.

Photometric redshifts and stellar masses were derived from the SEDs
using the \hypz\ code \citep{2000A&A...363..476B}, and fitting
photometric data with simple stellar populations (SSP) as well
as with composite populations described by $\tau$-models
(SFR$\propto e^{-t/\tau}$, with $\tau=$ from 0.1 Gyr to $\infty$),
and using the population models of \citet{2005MNRAS.362..799M}.
This parametrisation of the star formation history may be inappropriate for actively star forming galaxies (especially at high redshifts), because it assumes that all galaxies are caught at their minimum SFR. However, for passively evolving galaxies this assumption may not affect appreciably the redshift and mass determinations, as their SFR is indeed $\sim0$.

The 10 galaxies that we discarded as dusty contaminants are found
to be at $z_{\rm phot}<1.4$.
All 12 bona-fide ETGs have
photometric redshifts in the range of $1.4<z_{\rm phot}<1.7$, 
are fitted by SSP models \citep{2005MNRAS.362..799M} older than $\sim 1$ Gyr (i.e.,
with strong 4000\AA/Balmer breaks),
and have stellar masses in the range $2.5\times 10^{11}<M*<4\times
10^{11}M_\odot$, adopting the \citet{2003PASP..115..763C} IMF, as
reported in Table~\ref{tab:tab1}.  Following D05 and Maraston et al. (2006) we
expect that photometric redshifts of ETGs computed in this way should be accurate to 
within $\Delta z\lesssim0.1$.

\section{Surface brightness profile fitting and simulations} 
We used GALFIT \citep{2002AJ....124..266P} to fit 2D
\citet{1968adga.book.....S} profiles
to the
$I$-band (F814W filter) ACS images of the 12 ETGs.  The median
integration over the COSMOS field is 2028~s, the pixel-scale is $\sim
0\farcs05$, and the FWHM of the PSF is $\sim0\farcs097$.
GALFIT was run on sky-subtracted images, obtained by the \sex\
software \citep{1996A&AS..117..393B}.  Thus, the sky was fixed to a null value in the GALFIT procedure, and the galaxy S\'ersic index ($n$), effective radius (\re), and total magnitude ($m_{814}$) were left as free parameters. Other free parameters in the fit are the object position ($x_{\rm c}$, $y_{\rm c}$),  the position angle (PA) of the major axis $a$, and the axis ratio ($b/a$) between minor and major axis. On the other hand, leaving both $n$ and the sky background as free parameters in GALFIT may lead to unreliable results for galaxies with intrinsically high values of $n$ (see the GALFIT home page\footnote{http://users.ociw.edu/peng/work/galfit/TFAQ.html\#sensitivity}).
For each galaxy the adopted PSF was obtained from the closest suitable bright unsaturated star. 
We verified that swapping PSF by using different stars in the COSMOS
field or synthetic PSFs, built by means of the Tiny Tim\footnote{http://www.stsci.edu/software/tinytim software package \citep[v. 6.3][]{2004KristandHook}} gives virtually identical results.
When galaxy isophotes partially overlap with bright close neighbours, we fitted
galaxy and neighbours simultaneously (this was the case for objects
\#136738, \#157677, \#200390, \#254025). For smaller and fainter
neighbours we used masked regions.  Weight maps were used accounting for
shot noise and sky noise.

\begin{figure}
\begin{center}
\includegraphics[width=\columnwidth]{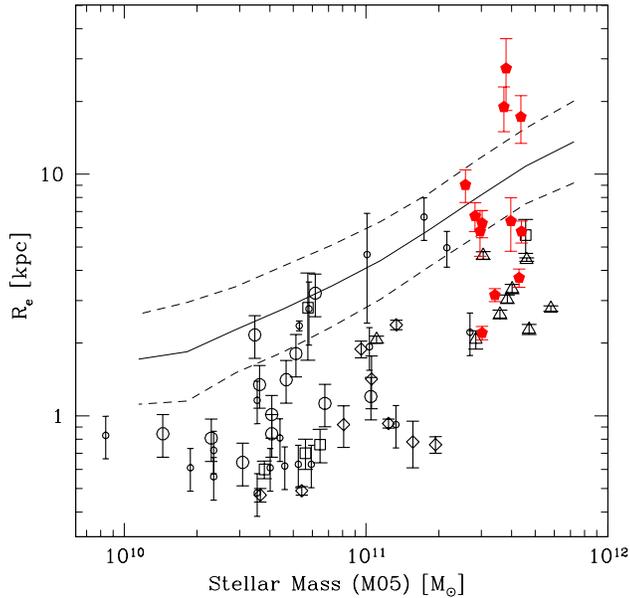}
\caption{Mass vs size for ETGs at $z\geq1.4$, compared with the local relation
  (solid line) and its 1~$\sigma$ dispersion (dashed lines). The figure is
  adapted from Figure 15 of \citet{2008A&A...482...21C} by adding our 12 ETGs, shown as red large filled pentagons. The open symbols represent a compilation of the literature results for passive galaxies: big open circles are from \citet{2008A&A...482...21C} (GMASS-CDFS); open triangles from \citet{2007MNRAS.374..614L} and \citet{2006ApJ...650...18T}; open square are the HUDF passive-BzK of Daddi et al. (2005) and Maraston et al. (2006); open diamonds show the sample of van Dokkum et al. (2008) after a mass-rescaling of a factor $\sim 1.4$, to  Maraston et al. (2005) models, for homogeneity with the other galaxies in the figure. The small open circles are from \citet{2007ApJ...656...66Z} (FIRES-HDFS), \citet{2007ApJ...671..285T} (FIRES- MS1054), \citet{2008ApJ...682..303M}, and \citet{2004Natur.430..184C} (K20 survey).}\label{fig:fig4} 
\end{center}
\end{figure}

 In Fig.~\ref{fig:fig2} we show the galaxy ACS I-band images, the corresponding GALFIT best fit model, and the residuals for each object.
The best-fit parameters $m_{814}$, the S\'ersic index $n$ (which measures the light central
concentration), and the half light radius  $\re$ (i.e. circularised radius of the ellipse
containing
half of the total luminosity of the best-fitting S\'ersic profile) are
reported in Table~\ref{tab:tab1}. 
As shown in Fig.~\ref{fig:fig4} and Table 1, only 3 of the 12 objects have very small effective radii
for their mass ($0.25<\re <0.48$ arcsec, i.e.,  $\sim 2$ to
$\sim 4$ kpc at $z\sim 1.5$). This qualifies them as `compact',
compared to the local ellipticals, whereas the remaining 9 objects
appear to be consistent with local ETGs, although
the effective radii of some of the objects with the largest size (e.g., \#136738, \#535544, \#157677) have also very large intrinsic uncertainties.
For all the objects but one we found $n>3$, as expected for typical ETGs. The exception is
object \#307881, one of the smallest in the sample, which may consist
of a bulge plus an exponential disk. On the contrary, the highest S\'ersic index ($n=7.29\pm 0.58$) is found for object \#136738. A large $n$ generally indicates an highly centrally concentrated light profile, compatible with such massive elliptical galaxy. However, for this object the presence of some structure at the centre of the residual image (Fig.~\ref{fig:fig2}) suggests that a central unresolved point-source (probably an AGN) may have contributed to increase the $n$ value.
Thus, we fitted object \#136738 by adding a central point-source, finding \re=1.22~arcs ($\sim$ 10 kpc at $z=1.6$), $n\sim4$ and $m_{814}$=22.28 for the S\'ersic component, and a magnitude of $m_{814}$=25.88 for the point source component. We conclude that the possible presence of a point-source would decrease by almost a factor of 2 the resulting $n$ value, but it would not dramatically affect the large \re ($\geq 10$ kpc) found for this object, still consistent with the size of similarly massive local ETGs.       

Fig.~\ref{fig:fig4} shows that our 12 ETGs span a large range in size,
although they have very similar masses. A large scatter in the size distribution was also found in other studies for less massive galaxies. 
Note that the 3 smallest objects in
our sample, with $\re\lesssim 4$ kpc, are among the faintest in the full
sample of 12 ETGs, and are all fainter than $m_{814}=22.9$.  This leads
us to wonder if the 3 small galaxies are really small as we have
measured, or if their apparent small $\re$ is due to an observational
bias. Thus, what we would like to clarify is whether the recovered
distribution of $n$ and $\re$ for our $z>1.4$ ETG sample, including
the 3 apparently small galaxies, could still be consistent with the
local distribution of these structural parameters.

In order to answer this question we have constructed a simulated set
of galaxies all having $\re=7$~kpc and $n=6$, which are the typical
values for local ETGs with stellar masses as large as those estimated
for the 12 ETG objects at $z>1.4$
\citep{2003MNRAS.343..978S,1993MNRAS.265.1013C}.  The GALFIT package
was used to generate synthetic galaxies with the chosen light profile
and having total fluxes similar to those of our ETG sample.  The
simulated galaxies were placed in empty sky regions of the COSMOS ACS
image, close to our real galaxies (within $\sim 1$ arcmin), resulting
in similar values of background noise. The noise and PSF images were
constructed as for the real observed galaxies, and we tested that the
results were not substantially altered if we used different PSF stars
for creating and fitting the simulated galaxies.

\begin{figure}
\begin{center}
\includegraphics[width=1.1\columnwidth]{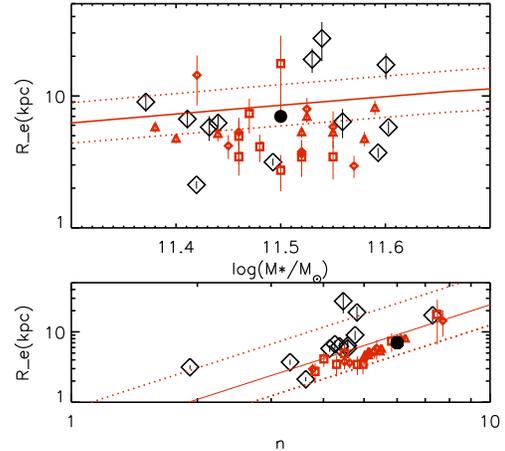}
\caption{In the top panel the stellar mass-size relation for our final
  sample of 12 ETGs (big open diamonds) is compared with the results obtained for 24 simulated objects (filled red symbols), all characterised by input S\'ersic parameters of n$_{\rm in}$=6, $R_{\rm e,in}\sim 7$~kpc and magnitude m$_{814,\rm in}$=22 (red triangles), 23 (red diamonds), or 23.5 (red squares). The bottom panel shows the S\'ersic index (n) - effective radius ($\re$) relation. The symbols are the same of the top panel.  
In both of the diagrams the local M*--$\re$ and n--$\re$ relations with their 1~$\sigma$ dispersions, from \citet{2003MNRAS.343..978S} and \citet{1993MNRAS.265.1013C}, respectively, are represented as red solid and dashed lines. The black filled circle shows the input n and $\re$ values common to all the 24 simulated objects. Note the large scatter in measured effective radius and S\'ersic index for the simulated objects, well reproducing the scatter observed for the real galaxies.}\label{fig:fig7}
\end{center}
\end{figure}

In the two panels of Fig.~\ref{fig:fig7} the local average $\re-M*$ and $n-\re$
relations and their 1~$\sigma$ dispersion are compared with $\re$,
$n$, and masses of our 12 ETGs, along with those obtained for 24
simulated objects, assuming different input luminosities, namely
$m_{814,\rm in}$=22,\ 23,\ 23.5,\ and positions on the frame.

It is worth emphasising that all the 24 simulated galaxies have the
same intrinsic effective radius and S\'ersic index (i.e., we are
neglecting here intrinsic size and $n$ scatter), and yet their output
$R_{\rm e,out}$ and $n_{\rm out}$ well reproduce the large scatter in
$\re$ and $n$ observed for our real sample.  The spread in size for
the simulated galaxies is a result of the spread in luminosity and of
the different local background noise realisations, with a tendency to
underestimate the size by up to a factor $\sim 3$ for the faintest
objects, with $m_{814,\rm in}=23-23.5$.  This suggests that the apparent
compactness of the 3 smallest objects (that are also among the faintest ones) may be just a consequence of their
lower S/N ratio, but intrinsically they could not be exceptionally
small in size compared to local ETGs of the same mass.  Thus, within our
sample of 12 very high mass ETGs at $<z_{\rm phot}>\sim 1.6$ we do not
find convincing evidence for the {\it small size syndrome} affecting
other samples of high-$z$ ETGs.  We conclude that the sample of 12
ETGs at $z>1.4$ shows a distribution of sizes that is fully consistent
with those of local $z=0$ ETGs of the same mass.

\begin{table*}
\begin{minipage}{165mm}
\caption{Characteristics of the final sample of 12 ETGs.}\label{tab:tab1}
\begin{scriptsize}
\begin{tabular}{cccccccccccc}
\#ID  & RA(J200) & Dec(J200) & m$_{814}$ & \re (arcs) &\re (kpc)  &  n  & z$_{ph}$ & log(M*) & age & A$_V$ &$K_s$\\  
\hline
 93409&   10:01:27.0044  &01:39:53.1198   &22.73$\pm$0.08 &0.78$\pm$0.11 &  6.70$\pm$0.92 &  4.26$\pm$0.34& 1.52&11.45 & 2.40 & 0.08  & 17.68\\
 112083&  09:59:11.4551  &01:42:25.2173   &22.93$\pm$0.05 &0.44$\pm$0.04 & 3.73$\pm$0.32 &  3.33$\pm$ 0.21& 1.61&11.63 & 2.75 & 0.00  & 17.51\\
 125656&  10:00:22.9577  &01:44:16.8353   &22.53$\pm$0.07 &0.73$\pm$0.09 & 6.26$\pm$0.81 &  4.37$\pm$ 0.35& 1.43&11.48 & 2.75 & 0.08  & 17.66\\
 136738&  09:59:47.0280  &01:45:52.2360   &21.98$\pm$0.11 &2.02$\pm$0.45 & 17.27$\pm$3.84 &  7.29$\pm$ 0.58& 1.59&11.64& 2.40 & 0.00  & 17.35\\
 157677&  10:01:34.2920  &01:48:32.0022   &22.51$\pm$0.20 &3.20$\pm$1.06 & 27.36$\pm$9.03 &  4.46$\pm$ 0.62& 1.70&11.58& 2.50 & 0.00  & 17.64\\
 200390&  10:01:01.9974  &01:54:32.4251   &23.43$\pm$0.11 &0.68$\pm$0.14 & 5.81$\pm$1.24 &  4.57$\pm$ 0.60& 1.61&11.47 & 2.00 & 0.08  & 17.68\\
 217431&  10:02:39.5288  &01:56:59.1162   &21.85$\pm$0.12 &2.22$\pm$0.47 & 18.92$\pm$3.98 &  4.81$\pm$ 0.44& 1.61&11.57& 2.40 & 0.00  & 17.56\\
 254025&  10:02:28.4912  &02:02:13.6899   &22.50$\pm$0.06 &0.68$\pm$0.07 & 5.79$\pm$0.61 &  4.14$\pm$ 0.26& 1.71&11.64 & 2.10 & 0.08  & 17.54\\
 307881&  10:02:35.6396  &02:09:14.3640   &22.95$\pm$0.04 &0.37$\pm$0.02 & 3.16$\pm$0.20 &  1.92$\pm$ 0.12& 1.52&11.53 & 2.50 & 0.08  & 17.59\\
 399354&  10:02:49.4055  &02:21:47.4012   &22.98$\pm$0.14 &0.75$\pm$0.20 & 6.39$\pm$1.60 &  4.57$\pm$ 0.64& 1.70&11.59 & 2.30 & 0.08  & 17.63\\
 472257&  10:01:06.0754  &02:31:35.1966   &23.04$\pm$0.04 &0.25$\pm$0.02 & 2.13$\pm$0.14 &  3.63$\pm$ 0.23& 1.52&11.46 & 2.40 & 0.00  & 17.69\\
 535544&  10:00:00.6592  &02:40:29.6220   &22.29$\pm$0.09 &1.06$\pm$0.16 & 9.03$\pm$1.39 &  4.76$\pm$ 0.35& 1.41&11.41 & 2.30 & 0.16  & 17.54\\
\hline
\end{tabular}
\end{scriptsize}
\begin{footnotesize}
From left to right: {\bf \#ID}: identification number in the catalogue of McCracken et al (2009); {\bf RA(J2000)} and {\bf Dec(J2000)}: coordinates from
WIRCAM/$K$-band data set; {\bf m$_{814}$}: ACS F814W $I$--band magnitude; {\bf \re}: effective radius in arcseconds and kpc; {\bf n}: S\'ersic index;
{\bf z$_{ph}$}: photometric redshifts, {\bf log(M*)}: galaxy stellar masses in M$_{\odot}$ units; mean stellar {\bf age} in Gyrs; {\bf A$_V$} reddening
parameter; {\bf $K_s$}: $K$-band WIRCAM+CHFT magnitude in Vega system). 
\end{footnotesize}
\end{minipage}
\end{table*}

\begin{figure}
\begin{center}
\includegraphics[width=1.0\columnwidth]{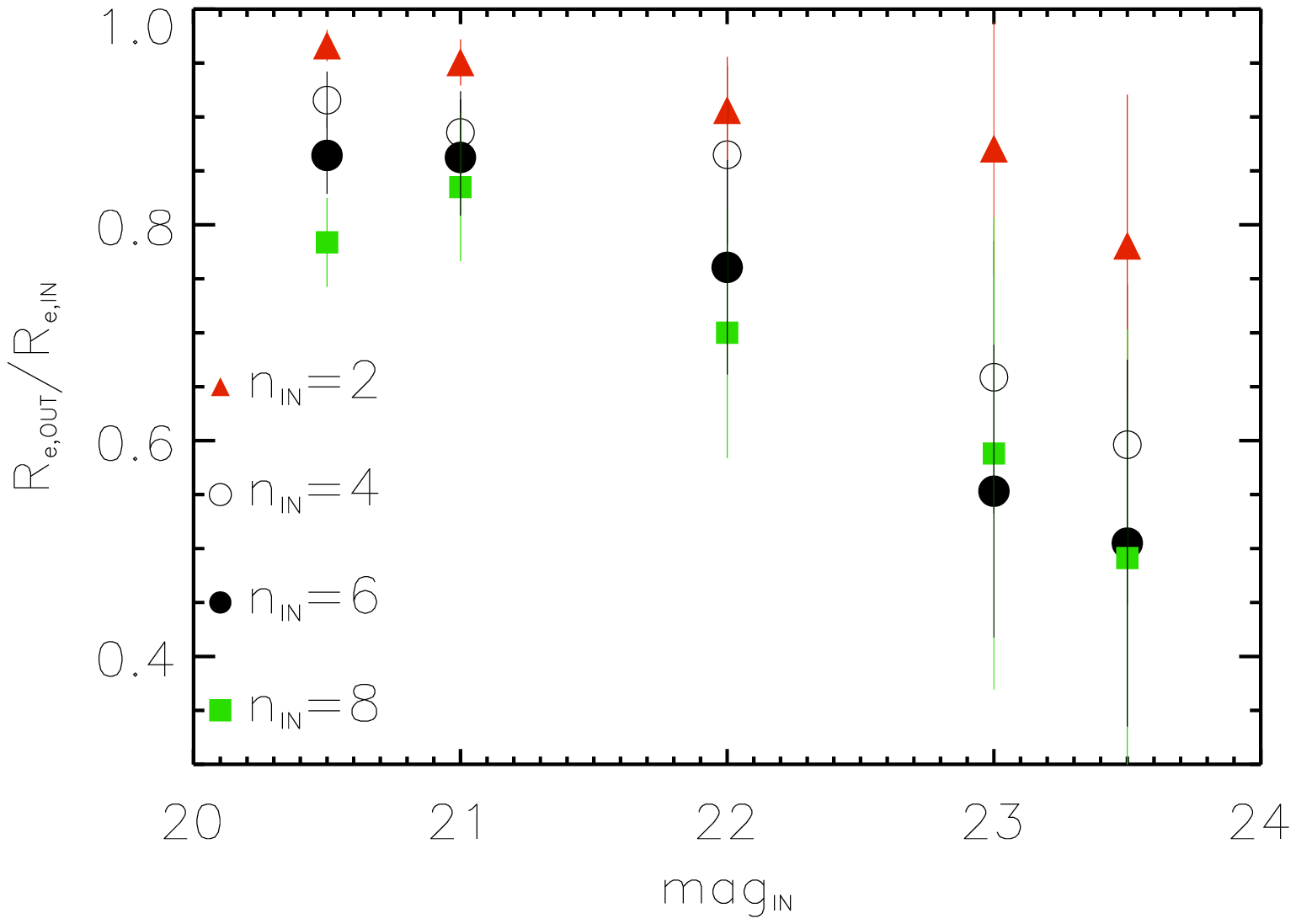}\\
\caption{Simulated galaxies. $R_{\rm e,out}/R_{\rm e,in}$ (kpc) as a function
  of input magnitude ($m_{814,\rm in}$=20.5, 21, 22, 23, 23.5), for
  a fixed large value of effective radius, i.e. $R_{\rm e,in} \sim0.8$~arcsec,
  corresponding to $R_{\rm e,in}\sim 7$~kpc at $z\sim 1.5$, and different values
  of input S\'ersic index, as labelled in the diagram. Worth noticing that the
  $R_{\rm e,out}$ estimate decreases with decreasing luminosity and increasing S\'ersic
  index of the source.}\label{fig:fig6}
\end{center}
\end{figure}

In order to understand these results and place them in the contest of 
previous findings we run a more general suite of simulations.
We built 1200
synthetic galaxies with  a much wider range of parameters:
$\re$=0.2, 0.4, 0.8, 1.6, and 3.2 arcsec
(corresponding to $\sim 2-30$~kpc at $z\sim 1.5$); $n$=2, 4, 6, and 8, 
and $m_{814}$=20.5, 21, 22, 23, and 23.5. In particular, since
the ACS/F814W magnitudes of the galaxies in our sample are in the
range $m_{814}\sim 22-23.5$, this allowed us to explore the
size-estimate dependence on the signal-to-noise ratio (S/N) over a
range of S/N$\simeq 50-200$ (here as well as below we refer to the
galaxy S/N as measured within a $1''$ diameter aperture). 

\begin{figure}
\begin{center}
\includegraphics[width=1.0\columnwidth]{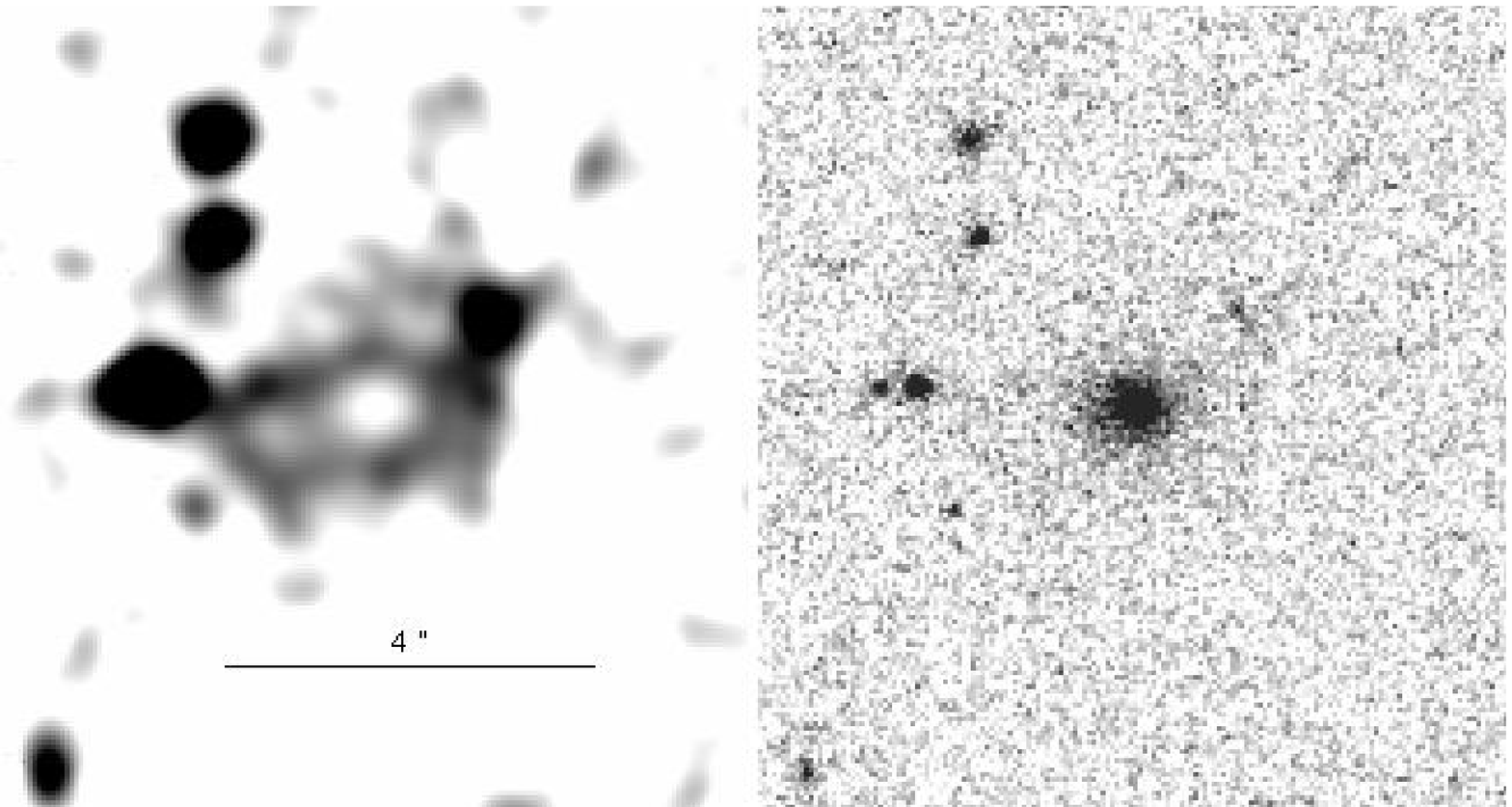}
\caption{
{\it Left:} the ACS I-band image of \#136738 (see also Fig.~\ref{fig:fig2}), the most massive and $K$-band
brightest ETG in our sample. We have masked the central bright core
and convolved the image with  a FWHM$=0\farcs5$ Gaussian kernel in order to enhance low surface 
brightness details. A very extended low surface brightness halo is clearly detected, surrounding the galaxy.
{\it Right:} the original ACS I-band image without smoothing.
  }
  \label{fig:halo}
\end{center}
\end{figure}

These simulations show that the size estimates is affected by a combination 
of three factors: i) the S/N of the galaxy image, ii) the intrinsic galaxy 
size (\re), and iii) the profile shape ($n$). The effects of these three factors are discussed below:

i) As expected, the brighter the simulated objects, i.e. the higher the S/N,
the better the accuracy in the recovered parameters. As illustrated in Fig.~\ref{fig:fig6}, the fainter the input magnitude, the smaller the recovered effective radius, i.e., the stronger the underestimate of the size.

ii) For simulated galaxies with parameters and S/N similar to those of
our 12 real ETGs ($m_{814}\simeq 22-23.5$), the mean difference between the
input and output effective radii turns out to be
$<\Delta\re/\re>=<(R_{\rm e,in}-R_{\rm e,out})/R_{\rm
e,in}>\simeq 0.15$, with a dispersion of
$\sigma(\Delta\re/\re)\sim0.46$, with most of this systematic offset
and large dispersion being due to the largest simulated objects. For
$R_{\rm e,in}<0''.8$ the average offset is very close to zero,
with a small scatter. Beyond $R_{\rm e,in}=0\farcs8$ we
found that the output $\re$ underestimates the input radius and the
scatter increases.

iii) The size estimates depend on the capability of recovering the
S\'ersic index, since the best fit parameters are correlated via the
$n-\re$ and $n-M*$ relations
\citep[]{1993MNRAS.265.1013C,1994MNRAS.271..523D}. Thus, for input
$n>4$ the output $n$ values tend to be underestimated up by a factor
of $\sim 1.5$, and mostly for galaxies with the lowest S/N.  Hence,
this underestimate of $n$ is accompanied by a size underestimate, as
illustrated in Fig~\ref{fig:fig6}. 
The recovered effective radius decreases with increasing $n$ index (and decreasing luminosity). Similar trends are
generally found for simulated galaxies with $R_{\rm e,in}\gtrsim
3.5$~kpc.

These experiments confirm that for objects with large effective radius
and S\'ersic index, as ETGs with masses of $>2.5\times
10^{11}$~M$_{\odot}$ are expected to be, with the typical S/N of our 12 ETGs 
(i.e. S/N$\leq50-100$) one could
still substantially underestimate $n$ and $\re$. This may be 
due to the fact that
the halos in high S\'ersic-index profiles are very extended
and therefore have very low surface brightness, so that they can be easily 
missed in the noise at high redshift  given that the 
surface brightness scales as $(1+z)^4$.

A remarkable exemplification of this situation is shown in Fig.~\ref{fig:halo}
for the galaxy \#136738, the most massive and K-band brightest ETG in our sample. 
A giant low-surface brightness halo is clearly detected, extending over some 4$''$
in diameter around the galaxy (about 34 proper kpc at $z=1.5$), at a typical 
level of the order of 26~mag~arcsec$^{-2}$. Given the faintness of this structure,
it would be obviously very easy to miss in the noise similar halos at 
a lower S/N ratio.

\section{Discussion and Conclusions}

Over the 2 square degrees COSMOS field, we have identified and studied
the 12 $K$-band brightest and most massive
($M>2.5\times10^{11}M_\odot$) ETGs at $z\gtrsim 1.4$. Out of them, 9
follow the same size-mass relation of local ellipticals,
and 3 appear to have smaller effective radii. However, our simulations
show that their effective radius may have been underestimated by
a factor of 2--3 because of their lower S/N. Hence, for the whole
sample we find no compelling evidence of departures from the local
size-mass relation.

Most previous studies of high-$z$ ETGs were restricted to much smaller
areas, hence they may have missed the brightest/largest ETGs similar
to those reported here, as their surface number density is very low,
i.e., $\sim 6.3\pm1.8$ deg$^{-2}$ to the limiting magnitude of the
present sample. Using the redshift range $1.4<z<1.8$ to estimate the
cosmological volume, this corresponds to a space density of
$1.4\pm0.4\times10^{-6}$~Mpc$^{-3}$.  This is roughly 10\% of the
space density of local ETGs selected with $M>2.5\times
10^{11}M_\odot$. It is not clear if the different results of the
present work compared to previous studies of high-$z$ ETGs are due to
the galaxies in our sample being much more massive than those in most
previous works. Combining the results of our 12 massive ETGs with the
literature results --so to span a wide range of masses (2 dex)-- we
note that the departure of ETGs at $1.4\leq z \leq 2$ from the $z=0$
size-mass relation seems to be differential with mass, with
$\re(z=0)/\re(z)$ being on average close to 1 for the most massive
systems, and $\sim 2-4$ for lower masses.  This is reminiscent of the
general downsizing scenario, whereby massive ETGs formed their stars
and assembled their mass more rapidly and at higher redshifts compared
to the less massive ones. Thus, the present results may extend the 
downsizing effect to
the internal structure of ETGs, with the most massive ETGs
reaching their internal and dynamical structure earlier and faster than
lower mass ETGs at the same redshift.

However, we also argue that bias in the galaxy profile fitting could
significantly affect the size measurements, as low-surface brightness
halos could remain undetected in high redshift ellipticals if observed
with low S/N ratios (e.g., Fig.~\ref{fig:halo}). Based on redshifted
mock images of local systems, \citet{2009arXiv0903.2479H} also
concluded that there is a tendency to underestimate the galaxy size
which is more pronounced in the most massive galaxies, due to their
large $\re$ and $n$. This is in agreement with the results of our
simulations.  They also stressed that elliptical galaxy profiles are
not perfectly ``S\'ersic'', which would contribute to the size
underestimate at high redshift.

Thanks to the exceptional brightness of our 12 ETGs, their S/N ratios
in the single-orbit COSMOS $I$-band images are among the highest
reached by HST+ACS imaging of high-$z$ ETGs, and are comparable to
those obtained by \citet{2005ApJ...626..680D} for \pbzks\ in the
extremely deep HUDF images.  On average, their S/N ratio is also $\sim3$ times
higher than that of galaxies in the GMASS study, which
used GOODS-S $z$-band data \citep{2008A&A...482...21C}. Hence, it is
possible that also in these studies the effect of the noise has led to
underestimate the size of some object by a factor of 2 or 3.
\citet{2008A&A...482...21C} used simulated objects to check the
reliability of their derived effective radii, and estimated an average
scatter on the galaxy size measurements of $\sim 20\%$, with no strong
bias.  This suggests that a size underestimate is unlikely to account
for the small size obtained for the bulk of the GMASS high-$z$ ETGs.
Still, the parameter space ($n$, $\re$, S/N) explored in the
simulations of \citet{2008A&A...482...21C} may not fully include that
occupied by the real GMASS galaxies, hence we cannot exclude that also
in this study the size of some of the intrinsically largest galaxies
may have been underestimated by more than a 20\%.

On the other hand, some studies using $H$-band (F160W) NICMOS-2
imaging \citep[e.g.][]{2007MNRAS.374..614L, 2008ApJ...677L...5V} have
also reached high S/N ratios, thanks to them sampling longer rest-frame
wavelengths where the flux of ETGs is much stronger than in the
near-UV. Using the NICMOS Exposure Time Calculator (ETC), we estimated
that the S/N ratios of galaxies in \citet{2007MNRAS.374..614L} and
\citet{2008ApJ...677L...5V} were on average quite comparable, or sometimes
even slightly higher than the ones obtained for our COSMOS ETG sample.
The masses of ETGs in the former study
(open triangles in Fig.~\ref{fig:fig4}) are similar to those of our galaxies, and yet the derived  effective radii
are up to 3 times smaller than those of local
ellipticals with similar masses. Based on simulations, the
authors argued that their typical image resolution and S/N led to
underestimate by $\sim 25\%$ the effective radii of
galaxies with an intrinsic $n=4$ and radii in the range of
$\sim 1-4$~kpc, and took into account this result in their galaxy size estimate. However, they did not verify the robustness of their
fitting procedure in recovering the light profiles of galaxies with
$\re>4$~kpc and $n>4$, typical of such extremely massive ETGs
($M>2.5\times10^{11}$~M$_{\odot}$), which we found instead to be affected by
a substantial bias.

Similarly, \citet{2008ApJ...677L...5V} found their ETGs $\sim$5 times smaller compared to local ETGs of similar mass.
However, their stellar masses were estimated using the stellar
population models of \citet{2003MNRAS.344.1000B} that do not include
stars in their TP-AGB phase, and therefore stellar masses of ETGs at
$z\gtrsim 1.4$ might have been overestimated by a factor $\sim 1.5-2$
\citep{2006ApJ...652...85M}, while the TP-AGB effect is negligible for
local ETGs.  In the size-mass diagram, this mass re-scaling would
reduce by $\sim 3$ times the average offset in size with respect to
the local relation (as shown in Fig.~\ref{fig:fig4}).  Moreover, allowing for a systematic underestimate
of the size of the galaxies with lower S/N, similar to the one found by Longhetti et al. (2007) for lower-redshift ETGs in NIC2 images, we conclude that some galaxy in the study of
\citet{2008ApJ...677L...5V} may actually follow the local size-mass
relation.  

\citet{2009MNRAS.392..718S} collected 32 high-redshift ETGs
with archival NICMOS data, among which 13 have given size comparable
to the local galaxies and the others are more than 1$\sigma$ smaller
compared to the local relation. Also in this case, based on the above arguments, we cannot exclude that the size of the galaxies with lower S/N and lower resolution (those observed only with NICMOS-3) may have been underestimated.
However, the authors suggested an age-dependent evolution, with the {\it young ETGs (mean stellar age $< 2$~Gyr) being more similar to their local counterparts, and the {\it old ETGs} being denser. Our results cannot confirm such claim, as our best-fit ages are all confined within a very narrow range (see Table~\ref{tab:tab1}).}

A potentially powerful test of the reliability of the size estimates
of high-$z$ ETGs consists in measuring their velocity dispersion
$\sigma_{\star}$, which would allow us to check for the consistency of
the three quantities entering the virial relation: $M_{\rm
dyn}\simeq5\times\re\sigma_{\star}^2/G$, with the constraint of the
stellar mass being less than the dynamical mass.
Due to the limits of the present-day telescopes, the low S/N of
the available spectra of ETGs at $z>1.4$  allowed only a few attempts
in this direction, most of which based on the analysis of average
stacked spectra of several objects \citep{2009ApJ...696L..43C,2009arXiv0906.3648C}. Both these attempts are based on the GMASS
sample of spectra \citep{2008A&A...482...21C}, and within the errors found
agreement between stellar and dynamical masses. However, \citet{2009arXiv0906.3648C} found a trend to underestimate the virial masses ($M_{\rm
dyn}$) on average by $\sim 30$\%, that may be ascribed to a size underestimate by a similar quantity. 
Thanks to a $>7$ S/N ratio of their spectra, Cappellari et al. also
measured the $\sigma_{\star}$ from the individual spectra of two of
the galaxies with the largest size in the GMASS sample (whose sizes are
consistent with those of local ETGs). The resulting agreement with the
masses and sizes previously estimated by Cimatti et al. (2008)
confirmed that these two objects are truly similar to local ETGs, as most of the ETGs in our sample. Most recently, \citet{2009arXiv0906.2778V} analysed the near-IR spectrum of an ETG at $z=2.2$ finding
a very high velocity dispersion ($\sigma_{\star}\sim 510$ km/s),
consistent with the small size measured for this object, which then
suggests a strong structural/dynamical evolution from $z\sim 2$ to 0
for similarly massive ETGs. Yet, the relatively low S/N of the near-IR
spectrum
implies a fairly large uncertainty in the measured $\sigma_{\star}$.
Recently some among us were able to obtain a good S/N near-IR spectrum of one of our 12 ETGs (object \# 254025) with the MOIRCS spectrograph at the SUBARU telescope, and a measurement of $\sigma_{\star}$ will be attempted (M. Onodera et al. in preparation).

We conclude that the fraction of high-redshift ETGs which are
genuinely undersized with respect to local ETGs is still poorly known.
Several massive ETGs at high redshift (certainly the majority in our
sample) are as large as their local counterparts, and therefore
recipes aimed at inflating putative small ETGs should avoid to create
oversize ETGs at low redshifts out of those having already reached
their final size at $z\gtrsim 1.4$. We also hint at a possible downsizing effect not only on stellar ages and mass assembly, but also on the structural properties of these galaxies, with the most massive ETGs being in all respects the first to reach their final configuration.

\section*{Acknowledgements}
We wish to thank Swara Ravindranath and Paolo Cassata for useful
discussions. This work has been supported by grants ASI/COFIS/WP3110+WP3400 I/026/07/0.
CM, ED, FS, HJMCC, and MO acknowledge funding support of French ANR under contracts
ANR-07-BLAN-0228 and ANR-08-JCJC-0008, and CM thanks CEA Saclay for 
hospitality and funding at the beginning of this work.

\bibliography{CM_v1}

\label{lastpage}
\end{document}